\documentclass{jfm}
\usepackage{graphicx}
\usepackage{epstopdf,epsfig}

\usepackage{amsmath}
\usepackage{amssymb}
\usepackage{subfig}
\usepackage{tikz,siunitx}
\usepackage{xcolor}

\usepackage{scalerel}

\newcommand{\solidline}{\protect\tikz[baseline]{\protect\draw[solid,line width=1pt](0.0mm,0.5ex)--(6.5mm,0.5ex)}}
\newcommand{\dashline}{\protect\tikz[baseline]{\protect\draw[dashed,line width=1pt](0.0mm,0.5ex)--(6.5mm,0.5ex)}}
\newcommand{\dotline}{\protect\tikz[baseline]{\protect\draw[dotted,line width=1pt](0.0mm,0.5ex)--(6.5mm,0.5ex)}}
\newcommand{\dotdashline}{\protect\tikz[baseline]{\protect\draw[dash dot,line width=1pt](0.0mm,0.5ex)--(6.5mm,0.5ex)}}

\newcommand{\circl}{\protect\tikz[baseline]{\protect\draw[solid,line width=0.4pt](2.mm,0.5ex) circle (0.7ex)}}

\newcommand{\tridown}{\protect\tikz[baseline]{\protect\draw[line width=0.4pt](0,1.6mm)--(1.84mm,1.6mm)--(0.92mm,0)--(0,1.6mm);}}
\newcommand{\squar}{\protect\tikz[baseline]{\protect\draw[line width=0.4pt](0,0) rectangle (1.6mm,1.6mm)}}

\shorttitle{Nonlinear mechanism of the SSP of wall-bounded flows}
\shortauthor{H. J. Bae, A. Lozano-Dur\'an and B. J. McKeon}

\title{Nonlinear mechanism of the self-sustaining process in the
buffer and logarithmic layer of wall-bounded flows}

\author{H. Jane Bae\aff{1}\corresp{\email{hjbae@caltech.edu}} \and A.
Lozano-Dur\'an\aff{2} \and Beverley J. McKeon\aff{1}}

\affiliation{\aff{1}Graduate Aerospace Laboratories, California
Institute of Technology, Pasadena, CA 91125, USA
\aff{2}Center for Turbulence Research, Stanford University, Stanford,
CA 94305, USA}

\begin{document}
\maketitle

\begin{abstract}
The nonlinear mechanism in the self-sustaining process (SSP) of
wall-bounded turbulence is investigated. Resolvent analysis is used to
identify the principal forcing mode which produces the maximum
amplification of the velocities in numerical simulations
of the minimal channel for the buffer layer and a modified
logarithmic (log) layer. The wavenumbers targeted in this study are
those of the fundamental mode that is infinitely long in the
streamwise direction and once periodic in the spanwise direction. The
identified mode is then projected out from the nonlinear term of the
Navier-Stokes equations at each time step from the simulation of the
corresponding minimal channel. The results show that the removal of
the principal forcing mode of the fundamental wavenumber can inhibit
turbulence in both the buffer and log layer, with the effect being
greater in the buffer layer.  Removing other modes instead of the
principal mode of the fundamental wavenumber only marginally affects
the flow.  Closer inspection of the dyadic interactions in the nonlinear term
shows that contributions toward the principal forcing mode come from a
limited set of wavenumber interactions. Using conditional
averaging, the flow structures that are responsible for generating the
nonlinear interaction to self-sustain turbulence are identified as
spanwise rolls interacting with oblique streaks. This method, based on
the equations of motion, validates the
similarities in the SSP of the buffer and log layer, and characterises
the underlying quadratic interactions in the SSP of the minimal
channel.
\end{abstract}

%----------------------------------------------------------------%
\section{Introduction}
%----------------------------------------------------------------%
% Self sustained cycle with based on the interaction of coherent structures
% (streaks and rolls)
Since the first experiments by \citet{Klebanoff1962} and
\citet{Kline1967}, the coherent structure of near-wall turbulence has been
extensively investigated. In the vicinity of the wall, the flow is
found to be highly organised and can be comprehended as a collection
of recurrent patterns usually referred to as coherent structures or
eddies, consisting of streamwise rolls/vortices and low- and
high-speed streaks \citep{Kline1967,Smith1983,Blackwelder1979} that
are involved in a quasi-periodic regeneration cycle
\citep{Robinson1991,Panton2001,Adrian2007}.  However, despite the
large efforts devoted to the subject, questions still remain 
regarding the exact mechanisms by which turbulence self-sustains
in wall-bounded turbulent shear flows and the dynamics in which these
structures interact.

\citet{Townsend1976} hypothesised that the log region would be
composed of self-similar energy-containing motions, the sizes of which
are proportional to their distance from the wall. By a suitable
superposition of these hypothetical motions, termed `attached eddies',
under the constraint of a constant Reynolds shear stress typical of
the log layer, Townsend predicted that the wall-parallel velocity
components of turbulence intensities in the log region would
exhibit logarithmic wall-normal dependence.  Traditionally,
wall-attached eddies have been interpreted as statistical entities
\citep{Marusic2010,Smits2011}, but recent works suggest that they can
also be identified as instantaneous features of the flow
\citep{Jimenez2018}.  The methodologies to identify instantaneous
energy-eddies are diverse and frequently complementary, ranging from
the Fourier characterisation of the turbulent kinetic energy
\citep{Jimenez2013,Jimenez2015} to adaptive mode decomposition
\citep{Hellstrom2016,Cheng2019,Agostini2019}, and three-dimensional
clustering techniques
\citep{delAlamo2006,Lozano-Duran2012,Lozano-Duran2014,Hwang2018}, to
name a few. 

% Previous studies on SSC & mechanisms of lift up and breakdown
In the buffer layer, important progress was made in the early 1990s
using the `minimal flow unit' approach, which revealed that buffer
layer streaks can self-sustain even when motions at larger scales are
inhibited and that their existence, therefore, relies on an autonomous
process \citep{Jimenez1991}. \citet{Hamilton1995} utilised a similar
approach for Couette flow, where either certain velocity modes were
suppressed to remove streak formation or disturbances were added to
allow streak breakdown. \citet{Jimenez1999} further confirmed that
this near-wall process is independent of the flow in the log and outer
regions by showing the survival of the near-wall motions in the
absence of outer turbulence. The consensus from these studies, along
with many others that followed
\citep[e.g][]{Waleffe1997,Schoppa2002,Farrell2017}, is that the
streaks are significantly amplified by the quasi-streamwise vortices
via the lift-up effect \citep{Landahl1975}; the amplified streaks
subsequently undergo a rapid streamwise meandering motion, reminiscent
of streak instability or transient growth, which eventually results in
the breakdown of the streaks and regeneration of new quasi-streamwise
vortices \citep{Swearingen1987,Waleffe1995,Kawahara2003}.  The cycle
is restarted by the generation of new vortices from the perturbations
created by the disrupted streaks through nonlinear interactions.

A similar but more disorganised scenario is hypothesised to occur for
the larger wall-attached energy-eddies within the log layer
\citep{Flores2010,Hwang2011,Cossu2017,Lozano-Duran2020}. The existence
of a self-similar streak/roll structure in the log layer consistent
with Townsend’s attached-eddy model has been supported by the
numerical studies by
\citet{delAlamo2006,Flores2010,Hwang2011,Lozano-Duran2012,Lozano-Duran2014},
among others. A growing body of evidence also indicates that the
generation of the log-layer streaks has its origins in the linear
lift-up effect
\citep{Kim2000,delAlamo2006,Pujals2009,Hwang2010,Moarref2013,Alizard2015}
in conjunction with Orr's mechanism \citep{Orr1907,Jimenez2012}.
Regarding roll formation, several works have speculated that they are
the consequence of a sinuous secondary instability of the streaks that
collapse through a rapid meander until breakdown
\citep{Andersson2001,Park2011,Alizard2015,Cassinelli2017}, while
others advocate for transient growth \citep{Schoppa2002} or 
parametric instability of the
streamwise-averaged mean flow as the generating mechanism of the rolls
\citep{Farrell2016}.

The specific mechanism by which the coherent large-scale structures
self-sustain can be further investigated by looking for the existence
of invariant solutions of the Navier-Stokes equations
\citep{Waleffe1998,Waleffe2001,Waleffe2003}. The nonlinear
steady-state solutions and travelling wave solutions in plane Couette
and the travelling wave solutions in plane Poiseuille flows, also
known as exact coherent structures (ECS), are a combination of the
three flow structures--streamwise vortices, streaks, and waves such
that they maintain each other despite viscous decay. The idea was that
these ECS act as an organising centre for the turbulence quasi-cycle.
\citet{Kawahara2001} detected periodic orbits resembling the
time-dependent version of Waleffe's SSP that shares the full
regeneration cycle of the near-wall coherent structures in plane
Couette turbulence. Further study was done in the large-scale motions
\citep{Cossu2017} using over-damped large-eddy simulation to isolate
the large-scale motions. However, while realistic turbulence does show
similarity to ECS solutions, the actual dynamics of wall-bounded
turbulence exhibits more complexity  than the ones observed in ECS.

Several mechanisms for vortex regeneration have been
proposed in the past, mostly focused on the buffer region,
\citep{Jimenez1991,Hamilton1995,Panton1997,Schoppa2002,Hwang2016} due
to the fundamentally nonlinear nature of the problem. Previous studies
successfully identified the dominant wavenumbers involved in the
nonlinear interaction \citep{Hamilton1995} and the structures
involved in the regeneration mechanism based on instantaneous
snapshots of turbulent flow
\citep{Jimenez1991,Hamilton1995,Schoppa2002}. However, additional
insight is necessary to pinpoint the exact nonlinear interaction
involved in the regeneration cycle, i.e., to understand the underlying
flow composition that leads to the meaningful nonlinear interaction
which produces the structures involved in the SSP, in realistic
turbulent flows. A meaningful addition to the previous studies would
be to find an orthonormal basis for the nonlinear term of the
Navier-Stokes equations that isolates the most important nonlinear
interaction leading to regeneration of streamwise vortices, which is 
the goal of the present study.

Various approaches rooted in the linearised equations 
have been utilised in the past to study turbulent flows such as 
rapid distorsion theory (RDT) of \citet{Batchelor1954} (see review by 
\citet{Hunt1990,Cambon1999} for more details), linear stability theory
\citep{Lin1944,Malkus1956,Reynolds1972}, optimal perturbation
\citep{Butler1993,Farrell1993}, input-output analysis 
\citep{Jovanovic2005}, and resolvent analysis 
\citep{McKeon2010,Mckeon2017}, among others. Early works focused their efforts 
on transition to turbulence while more recent studies have have been 
devoted to fully turbulent flows. Here we focus on the latter using 
resolvent analysis, as it has been successful at identifying the most 
energetic motions in actual turbulent flows by approximating the nonlinear
forcing from the interaction of highly amplified coherent structures 
\citep{McKeon2010,Mckeon2017}. Resolvent analysis identifies pairs of
response (velocity) and forcing (nonlinear) modes and the
corresponding amplification factor from the linearised Navier-Stokes
operator while taking into account the non-normality of the linear 
Navier-Stokes equations. This results in a set of orthonormal bases for 
the velocity and nonlinear terms. A more detailed comparison of previous 
works on linearised  Navier-Stokes equations with the resolvent analysis 
can be found in \citet{Mckeon2017}.

It has been shown that a rank-one approximation of the resolvent modes 
captures the characteristics of the most energetic modes of 
wall-bounded turbulent channels \citep{Moarref2013}. We postulate 
that the principal (most
amplified) forcing mode then must have the largest impact on the flow
and, in particular, the regeneration cycle, and we show that the
turbulence can be weakened by removing the nonlinear component
corresponding to the principal forcing mode in both the buffer and the
log layer. The orthonormality of the resolvent
modes allows the removal of the spatio-temporal structure 
corresponding to the principal forcing mode by extracting the 
projection of forcing mode onto the nonlinear term.
Furthermore, by identifying the most important contribution to the 
nonlinear term, conditional averaging can identify the underlying 
interaction that leads to the generation of this nonlinear term, 
making it possible to study the structures responsible for the nonlinear
mechanism in the SSP. A preliminary version of this work 
can be found in \citet{Bae2020}. 

The paper is organised as follows. We first introduce the method used
to identify and remove resolvent forcing modes from the nonlinear term
computed from the numerical simulations of a low- and 
moderate-Reynolds-number minimal
channel flow simulations, representing the buffer and log-layer
dynamics, in \S\ref{sec:methods}. We then present the resulting
changes in the flow statistics once the resolvent forcing modes are
projected out in \S\ref{sec:results}, where it is shown that the
removal of the principal forcing modes can significantly reduce
turbulence in both the buffer and log layers. The precursor velocity
structures involved in the generation of the most amplified nonlinear
term, and thus the structures involved in the nonlinear mechanisms of
the SSP, are identified in \S\ref{sec:nonlinear}. Finally, our
findings are summarised in \S\ref{sec:conclusions}.

%----------------------------------------------------------------%
\section{Methods}\label{sec:methods}
%----------------------------------------------------------------%
In the following, we consider a turbulent flow between two parallel
walls. The streamwise, wall-normal and spanwise directions are denoted
by $x$, $y$, and $z$, respectively. The streamwise and spanwise
directions are periodic. The flow velocities in the
corresponding directions are given by $\bar{U}$, $\bar{V}$, and
$\bar{W}$. We define two decompositions of the flow velocities
$\bar{U} = U + u = \bar{u} + u'$, where $U$ is the velocity averaged
over the homogeneous directions and time and $\bar{u}$ is the velocity
averaged over the homogeneous directions only (analogously defined 
for $\bar{V}$ and $\bar{W}$).  The flow is characterised by the 
friction Reynolds number $\Rey_\tau = \delta u_\tau/\nu$, where 
$\delta$ is the half channel height, $u_\tau$ is the friction velocity,
and $\nu$ is the kinematic viscosity.

%----------------------------------------------------------------%
\subsection{Principal forcing modes}\label{sec:methods:resolvent}
%----------------------------------------------------------------%

In order to identify the most amplified nonlinear component in the
turbulent channel flow, we perform the resolvent analysis with the
base mean flow being the mean velocity of a turbulent channel flow, 
$\boldsymbol{U}=(U,V,W)$.
The incompressible Navier-Stokes equations can be Fourier transformed
in homogeneous directions and time and reorganised as 
\begin{equation}
-\mathrm{i}\omega\tilde{\boldsymbol{u}} +
(\boldsymbol{U}\cdot\hat{\nabla})\tilde{\boldsymbol{u}} +
(\tilde{\boldsymbol{u}}\cdot\hat{\nabla})\boldsymbol{U} + 
\hat{\nabla}\tilde{p} -
\frac{1}{\Rey_\tau}\hat{\Delta}\tilde{\boldsymbol{u}} = 
\tilde{\boldsymbol{f}},\quad
\hat{\nabla}\cdot\tilde{\boldsymbol{u}} = 0,
\end{equation}
for each $(k_x,k_z,\omega)\neq(0,0,0)$, where
$(\tilde{\cdot})$ is the Fourier transform in time and space,
$\tilde{\boldsymbol{u}}(k_x,k_z,\omega) =
[\tilde{u},\tilde{v},\tilde{w}]^T$, $\boldsymbol{U}= [U,V,W]^T$,
$\tilde{\boldsymbol{f}}(k_x,k_z,\omega) =
[\tilde{f}_u,\tilde{f}_v,\tilde{f}_w]^T$ denotes the nonlinear
advection terms, and $\tilde{p}(k_x,k_z,\omega)$ is the pressure, all
of which are functions of wall-normal distance $y$. The operators
$\hat{\nabla} = [\mathrm{i}k_x,\partial_y,\mathrm{i}k_z]^T$ and
$\hat{\Delta} = \partial_{yy}-k_x^2-k_z^2$.  Here, the triplet
$(k_x,k_z,\omega)$ denotes the streamwise and spanwise wavenumbers and
the temporal frequency, respectively. The length, velocity, and time
scales are nondimensionalised using $\delta$, $u_\tau$, and
$\delta/u_\tau$, respectively, and $k_x$, $k_z$, and $\omega$ are
nondimensionalised using $\delta^{-1}$ and $u_\tau/\delta$,
respectively. Further, to facilitate the notation, we define
$k_x^\circ = 2\pi\delta k_x/L_x$ and $k_z^\circ = 2\pi\delta k_z/L_z$
such that integer values of $k_x^\circ$ and $k_z^\circ$ indicate the
number of wavelengths that fit in the streamwise and spanwise 
domain of size $(L_x,L_z)$, and 
for any function $\zeta(k_x,k_z,\omega)$,  $\zeta^{(a,b,c)} =
\zeta(k_x^\circ=a,k_z^\circ=b,\omega=c)$.

Equivalently, we can express the Navier-Stokes equations
for $(k_x,k_z,\omega)\neq(0,0,0)$ as 
\begin{equation} \left[\begin{array}{c}
\tilde{\boldsymbol{u}}(k_x,k_z,\omega)\\
\tilde{p}(k_x,k_z,\omega)\end{array}\right] =
\mathcal{H}(k_x,k_z,\omega)  \left[\begin{array}{c}
\tilde{\boldsymbol{f}}(k_x,k_z,\omega)\\ 0 \end{array}\right].
\label{eq:NS2}
\end{equation}
We refer to the linear operator $\mathcal{H}(k_x,k_z,\omega)$,
which takes $\boldsymbol{U}$ as input, as the
resolvent operator. The singular value decomposition of the resolvent
operator returns an ordered basis pair
$\{\tilde{\boldsymbol{\psi}}_j,\tilde{\boldsymbol{\phi}}_j\}$ along
with the associated singular value $\sigma_j$ ($\sigma_1 \ge \sigma_2
\ge \cdots \ge 0$) which can be used to express \eqref{eq:NS2} as
\begin{equation} 
\left[\begin{array}{c}
\tilde{\boldsymbol{u}}(k_x,k_z,\omega)\\
\tilde{p}(k_x,k_z,\omega)\end{array}\right] =
\sum_{j=1}^\infty\sigma_j(k_x,k_z,\omega)
\tilde{\boldsymbol{\psi}}_j(k_x,k_z,\omega) \left\langle
\tilde{\boldsymbol{\phi}}_j(k_x,k_z,\omega), \left[\begin{array}{c}
\tilde{\boldsymbol{f}}(k_x,k_z,\omega)\\ 0
\end{array}\right]\right\rangle, 
\end{equation}
where $\langle\cdot,\cdot\rangle$ is the inner product corresponding
to the kinetic energy norm, and the basis
$\tilde{\boldsymbol{\phi}}_i$ and $\tilde{\boldsymbol{\psi}}_i$ are
unitary.  We refer to $\tilde{\boldsymbol{\psi}}_j$ as the response
modes and $\tilde{\boldsymbol{\phi}}_j$ as the forcing modes. The
former identifies the most amplified coherent structures, which are
considered to contain most of the energy. The latter are the basis for
the nonlinear terms that create the response modes via the linear
resolvent operator. 

While the eddy viscosity formulation of the resolvent analysis, as in
\citet{Morra2019}, has been shown to improve the prediction of 
statistics for most energetic streamwise-constant streaks, a 
more recent study by \citet{Symon2020} shows that eddy viscosity 
does not respect the conservative nature of the nonlinear energy transfer,
which must sum to zero over all scales. Consequently, it is less 
effective for scales that receive energy from the nonlinear terms.
Thus, for the purposes of this study, we find the original resolvent 
formulation without the additional eddy-viscosity term more relevant.

Note that due to the symmetry in the channel flow, the resolvent modes
appear in pairs, and the modes corresponding to the two
largest singular values $\sigma_1$ and $\sigma_2$ can be decomposed
into two components that are symmetric about $y=1$. 
For the purpose of identifying modes
that act primarily on the top or bottom of the channel, we 
define the principal mode as the linear combination of the two
largest resolvent modes as follows. To find a principal forcing mode that acts
mostly on the bottom half of the channel ($0\le y\le 1$), we first
define the symmetric and anti-symmetric modes as a linear combination
that is symmetric or anti-symmetric, respectively, with respect to
$y=1$ and has positive real component when integrated from  $y=0$ to
$y=1$.  We then focus on the principal forcing mode
$\tilde{\boldsymbol{\phi}}_1 = [\tilde{\phi}_{1,u},
\tilde{\phi}_{1,v}, \tilde{\phi}_{1,w}, 0]^T$ computed by taking the
sum of the symmetric and anti-symmetric linear combination of the
pair, such that the sum of the two will mostly affect flow in the
bottom half ($0\le y \le 1$) of the channel.  Consequently,
$\tilde{\boldsymbol{\phi}}_2$ is the difference of the symmetric and
anti-symmetric linear combination of the pair corresponding to these
two largest singular values such that the projection of the resolvent
mode will mostly affect flow in the top half of the channel. All
subsequent modes $\tilde{\boldsymbol{\phi}}_{3,4,\cdots}$ are defined
analogously.

%----------------------------------------------------------------%
\subsection{Numerical simulation}\label{sec:methods:numerical}
%----------------------------------------------------------------%

% numerics
Two sets of numerical simulations are performed by solving the
incompressible Navier-Stokes equations
\begin{subequations}
\label{eq:incomp_NS}
\begin{equation}
\frac{\partial \bar{\boldsymbol{U}}}{\partial t} =  
-(\bar{\boldsymbol{U}}\cdot{\nabla})\bar{\boldsymbol{U}} - 
{\nabla}{p} + \frac{1}{\Rey_\tau}{\Delta}\bar{\boldsymbol{U}}, 
\quad
\nabla\cdot\bar{\boldsymbol{U}} = 0,
\tag{\theequation a,b}
\end{equation}
\end{subequations}
using a computational domain tailored to isolate the most energetic
eddies in either the buffer layer \citep{Jimenez1991} or the log layer
\citep{Bae2019b}. This can be considered the simplest numerical set-up
to study the SSP for wall-bounded energy-containing eddies of a given
size. 

In order to isolate the most energetic eddies in the buffer layer,
defined as $5<y^+<30$, where the superscript $+$ denotes viscous
units, we perform a direct numerical simulation (DNS) of an 
incompressible turbulent channel flow at
$\Rey_\tau \approx 186$. The simulations are performed by discretising
the incompressible Navier-Stokes equations with a staggered,
second-order accurate, central finite-difference method in space
\citep{Orlandi2000}, and an explicit third-order accurate Runge-Kutta
method for time advancement \citep{Wray1990}. The system of equations
is solved via an operator splitting approach \citep{Chorin1968}.
Periodic boundary conditions are imposed in the streamwise and
spanwise directions, and the no-slip condition is applied at the
walls.  The code has been validated in previous studies of turbulent
channel flows \citep{Bae2018a,Bae2019a}. The streamwise, wall-normal,
and spanwise domain sizes are $L_x^+ \approx 340$, $L_y^+\approx 372$
and $L_z^+ \approx 170$, respectively.  \citet{Jimenez1991} showed
that simulations in this domain constitute an elemental structural
unit containing a single streamwise streak and a pair of staggered
quasi-streamwise vortices, which reproduce reasonably well the
statistics of the flow in larger domains.  The grid spacings in the
streamwise and spanwise directions are uniform with $\Delta_x^+\approx
10.6$ and $\Delta_z^+\approx 5.3$; non-uniform meshes are used in the
wall-normal direction, with the grid stretched toward the wall
according to a hyperbolic tangent distribution with
$\min(\Delta_y^+)\approx 0.17$ and $\max(\Delta_y^+)\approx 7.6$. 

\begin{figure}
\begin{center}
\vspace{0.12cm}
\subfloat[]{\includegraphics[width=0.4\textwidth]{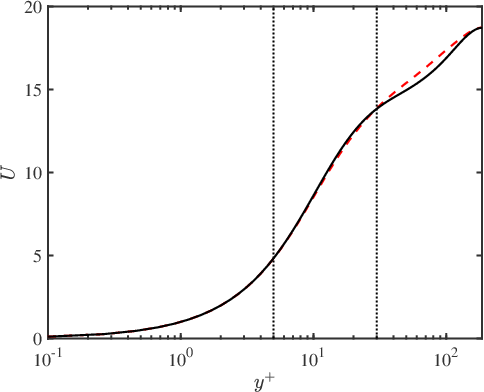}}
\hspace{0.3cm}
\subfloat[]{\includegraphics[width=0.4\textwidth]{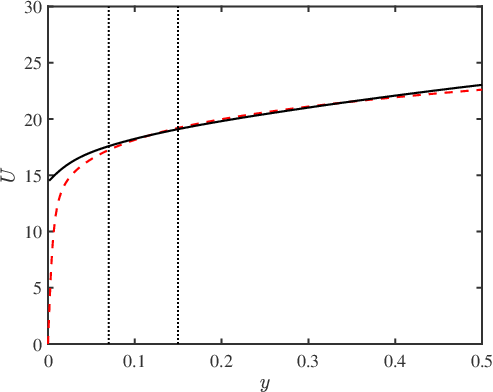}}\\
\subfloat[]{\includegraphics[width=0.4\textwidth]{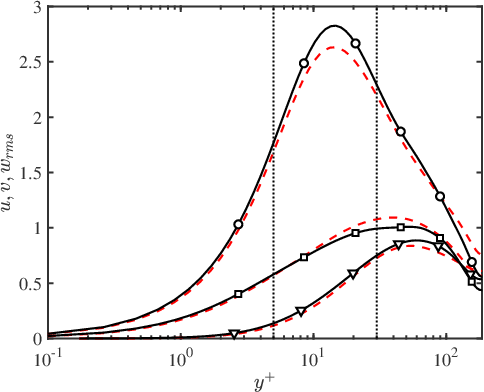}}
\hspace{0.3cm}
\subfloat[]{\includegraphics[width=0.4\textwidth]{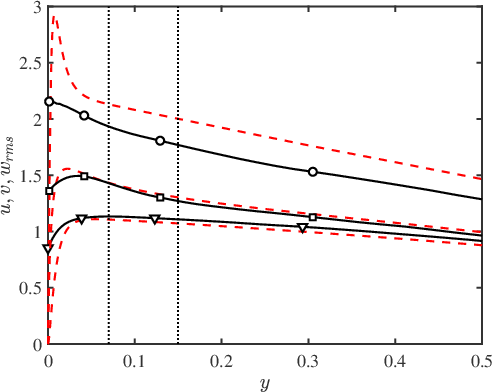}}\\
\subfloat[]{\includegraphics[width=0.4\textwidth]{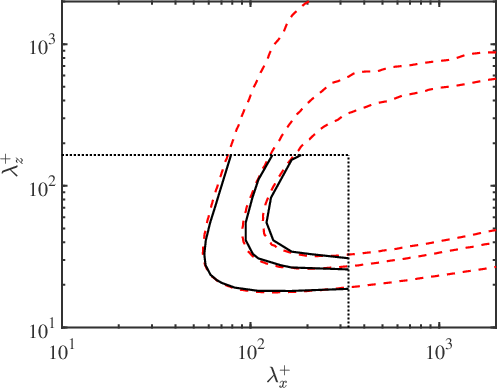}}
\hspace{0.3cm}
\subfloat[]{\includegraphics[width=0.4\textwidth]{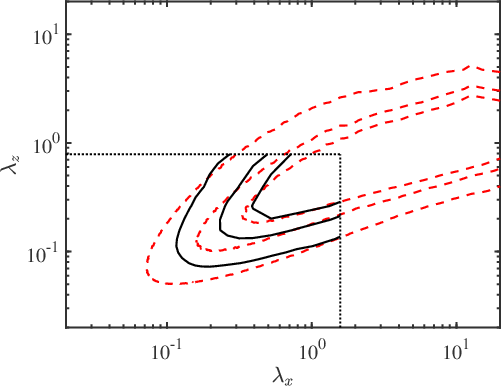}}\\
\caption{(a) Mean streamwise velocity profile (c) turbulence
intensities, and (e) energy spectra at $y^+ = 15$ of the buffer layer
minimal channel (black) compared to the mean velocity profile of
the channel flow for the domain size of $12\pi \times 2\times 4\pi$
(red) at $\Rey_\tau\approx 186$ from
\citet{delAlamo2003}. (b) Mean streamwise velocity profile, (d)
turbulence intensities, and (f) energy spectra at $y = 0.15$ of the
log layer minimal channel (black;\solidline) compared to the mean velocity
profile of the channel flow for the domain size of $8\pi \times
2\times 3\pi$ (red;\dashline) at $\Rey_\tau\approx 2003$ from
\citet{Hoyas2006}. The mean streamwise velocity profile for the
minimal channel in (b) is vertically shifted by 8.4 wall units such
that the velocity at $y = 0.15$ coincides with the larger-domain case.
Turbulence intensities shown for streamwise (\circl), wall-normal
(\tridown), and spanwise (\squar) components.  Dotted lines in (a--d)
indicate lower and upper domains for buffer and log layer for
corresponding cases. Dotted lines in (e,f) indicate the domain size of
the minimal channel. Contour levels are 1, 5, and 10\% for (e) and 10,
25, 40\% for (f) of the maximum value of the large-domain cases.
\label{fig:Umean}}
\end{center}
\end{figure}

In the case of the log layer, defined as $3\Rey_\tau^{-1/2}<y< 0.15$
\citep{Marusic2013}, we perform a large-eddy simulation (LES) of an
incompressible turbulent channel flow at $\Rey_\tau \approx 2003$ with
modified wall boundary conditions. The setup of the simulations are
similar to the buffer layer case, but the no-slip boundary condition
at the walls is replaced by a slip boundary condition of the form 
\begin{equation}
\bar{U} = l_s \frac{\partial \bar{U}}{\partial y},\quad 
\bar{V} = l_s \frac{\partial \bar{V}}{\partial y}, \quad 
\bar{W} = l_s \frac{\partial \bar{W}}{\partial y},
\end{equation}
where $l_s = 0.05$ is the slip length. The choice to use LES with the
slip boundary condition rather than DNS is deliberate in order to
remove the presence of small scale motions from the flow
\citep{Leonard1975,Hwang2016,Cossu2017,Bae2019b} and to suppress the
formation of near-wall viscous layers and buffer layer eddies
\citep{Lozano-Duran2016,Lozano-Duran2019a,Bae2019b}, as we are
interested in energy-containing motions of the log layer.
This modified channel flow is tailored to isolate the
large scale motions of the log layer. While the simulated log layer
produces healthy turbulence only in a limited range of wall-normal
locations, this allows a simplified approach to study the SSP of the
isolated scales.

The choice of the slip length is such that the adaptation length
\citep{Lozano-Duran2019a}, i.e., the vertical distance from the
boundary above which the flow recovers to the nominal no-slip flow
statistics, is below the lower bound of the log layer for this
Reynolds number. The anisotropic minimum dissipation model
\citep{Rozema2015} is used as the subgrid-scale model for the LES.
The streamwise, wall-normal, and spanwise domain sizes of the
simulation are $L_x \approx 1.57$, $L_y = 2$, and $L_z\approx 0.79$,
respectively.  This domain size corresponds to a minimal box
simulation for the log layer and is sufficient to isolate the relevant
dynamical structures involved in the bursting process
\citep{Flores2010}. Moreover, the choice of the domain size is such
that the wall-normal distance below which flow exhibits healthy
turbulence, $l_d \approx L_z/3 \approx 0.25$ \citep{Flores2010}, is
above the upper bound of the log layer.  The domain is discretised
using $N_x = 256$ and $N_z = 64$ points in the streamwise and spanwise
directions, and $N_y = 101$ in the wall-normal direction. The grid
spacings in the streamwise and spanwise directions are uniform with
$\Delta_x \approx 0.006$ and $\Delta_z \approx 0.012$ such that the
grid captures 90\% of the turbulent kinetic energy at $y = 0.15$,
which is necessary to accurately capture the coherent structures
present in the flow \citep{Lozano-Duran2019b}.  Non-uniform meshes are
used in the wall-normal direction, with the grid stretched toward the
wall according to a hyperbolic tangent distribution with
$\min(\Delta_y) \approx 0.003$ and $\max(\Delta_y) \approx 0.04$.
\citet{Bae2019b} showed that simulations in this domain with the
modified boundary conditions reproduce the statistics
in the log layer of the flow in larger domains while
suppressing the formation of the near-wall eddies.  For details
regarding this simulation, the reader is referred to \citet{Bae2019b}.

The flow is simulated for more than $100$ time units after transients
for computation of the mean streamwise velocity profile in the 
undisturbed minimal channel, which we label $U^{F}$,
shown in figure \ref{fig:Umean}(a,b) for the two cases. The resolvent
modes $\tilde{\boldsymbol{\phi}}_i$ and $\tilde{\boldsymbol{\psi}}_i$
for the minimal buffer and log layer were then computed with the 
respective $U^{F}$, using the
same staggered, second-order accurate, central finite difference
method in the wall-normal direction and the corresponding wall
boundary conditions. The Fourier discretisation in the computation of
the resolvent modes in the streamwise and spanwise directions was
updated to use the modified wavenumber corresponding to a staggered
second-order finite-difference method. The mismatch of the mean
velocity profile for the buffer layer case away from the wall
($y^+>30$) is expected due to the smaller box that is designed to
remove larger scales of motion and is in agreement with previous
studies utilising minimal channels \citep{Jimenez1991}. In the log
layer case, the simulation is designed to remove both the viscous
scales and subgrid scales as well as the larger outer region scales,
resulting in the mismatch in both the buffer and the outer region.
The disruption of the viscous scale can be better observed from the
lack of the inner peak in the streamwise turbulence intensity (figure
\ref{fig:Umean}(d)).

The new simulations are started from an initial condition from the
channel flow described above that is modified such that $\bar{u} =
U^{F}$. Then, an appropriate forcing at each time step is applied to
(\ref{eq:incomp_NS}a) to freeze the mean such that $\bar{u} = U^{F}$,
or equivalently, $\partial \bar{u}/\partial t = 0$. This is done by
setting the $(k_x,k_z)=(0,0)$ mode of the right-hand side of
(\ref{eq:incomp_NS}a) to zero at each time step. The mean velocity
profile is frozen so that the most amplified resolvent modes remain
constant throughout the simulation.  It is also consistent with the
aim of the study, which is to characterise the mechanisms that sustain
the fluctuating velocities generated by the actual turbulent mean
flow.  \citet{Tuerke2013} showed that turbulent channel flows with
prescribed correct mean velocity profiles result in naturally
occurring fluctuating velocities. An additional case without freezing
the mean, i.e., not applying the additional forcing to set
$\partial\bar{u}/\partial t = 0$, is performed later in
\S\ref{sec:results:180m} to study the effect of freezing the mean. 

The turbulence intensities and the energy spectra of the frozen-mean
minimal-domain simulations are compared to the large-domain
counterparts in figure \ref{fig:Umean}(c--f). We see that in the
regions bounded by the vertical dotted lines in figure
\ref{fig:Umean}(c,d), which signify the lower and upper bound of the
buffer and log layer in the respective cases, the turbulence
intensities follow the trend of its large-domain counterpart. The
underprediction of the turbulence intensities in the log-layer
simulation is due to the use of LES, where energy from the small and
largest scale motions are not included. The two-dimensional
premultiplied energy spectra at $y^+=15$ and $y = 0.15$, respectively,
are shown in figure \ref{fig:Umean}(e,f), and the results match well
with the large-domain counterparts, with the exception of the missing
large scales motions which extend beyond the respective domains and
the small scale contributions in the log-layer, which are modelled and
not resolved in LES.  

In order to study the effect of the forcing modes on the flow, a
separate simulation is advanced in time by removing the projection of
the nonlinear term, $\boldsymbol{f}' =
-(\boldsymbol{u}'\cdot\nabla)\boldsymbol{u}'$ onto the resolvent
forcing mode $\tilde{\boldsymbol{\phi}}_1$ for the chosen
$(k_x,k_z,\omega)$ at each time step. Note that $u' = u$ in the case
the mean is frozen in time.  This is done by first taking the Fourier
transform of the nonlinear term in the homogeneous directions and
computing the projection
\begin{equation} \label{eq:projection}
\hat{\boldsymbol{g}}_1(k_x,k_z,\omega) = 
\left\langle
\left[\begin{array}{c}
\hat{\boldsymbol{f}'}(k_x,k_z)\\
0
\end{array}\right],
\tilde{\boldsymbol{\phi}}_1(k_x,k_z,\omega)
\right\rangle 
\left[\begin{array}{c}
\tilde{\boldsymbol{\phi}}_{1,u}(k_x,k_z,\omega)\\
\tilde{\boldsymbol{\phi}}_{1,v}(k_x,k_z,\omega)\\
\tilde{\boldsymbol{\phi}}_{1,w}(k_x,k_z,\omega)
\end{array}\right]
\end{equation}
at each time step.  The projection is then removed by subtracting the
inverse Fourier transformed $\hat{\boldsymbol{g}}_1$ to the
right-hand-side of (\ref{eq:incomp_NS}a) and advancing in time.
Symmetry of the Fourier modes is preserved by also removing
$\hat{\boldsymbol{g}}_1^{*T}$, the conjugate of
$\hat{\boldsymbol{g}}_1$ (where superscript $*$ denotes conjugate
transpose and $T$ denotes transpose), from the nonlinear term
$\hat{\boldsymbol{f}'}(-k_x,-k_z)$ at each time step. Projections onto
$\tilde{\boldsymbol{\phi}}_i$ are analogously defined as
$\hat{\boldsymbol{g}}_i$. 

For the remainder of the paper, we denote the channel flow simulation
with the mean fixed at each time step but no forcing mode removed as
the {\it undamped} case and the simulation with the forcing mode
removed as the {\it damped} case.

%----------------------------------------------------------------%
\section{Effect of principal forcing modes on turbulence intensities}
\label{sec:results}
%----------------------------------------------------------------%

%----------------------------------------------------------------%
\subsection{Buffer layer}\label{sec:results:180m}
%----------------------------------------------------------------%

%
\begin{figure}
\begin{center}
\vspace{0.12cm}
\subfloat[]{\includegraphics[width=0.45\textwidth]{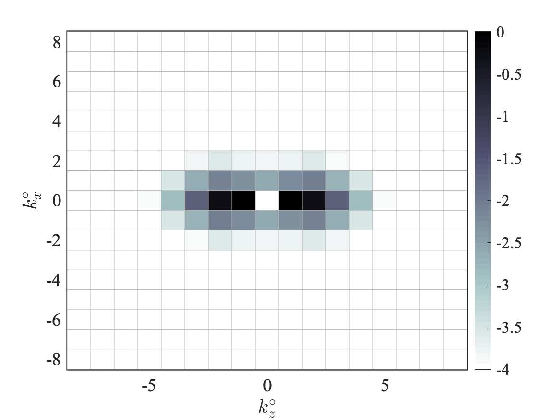}}
\hspace{0.3cm}
\subfloat[]{\includegraphics[width=0.40\textwidth]{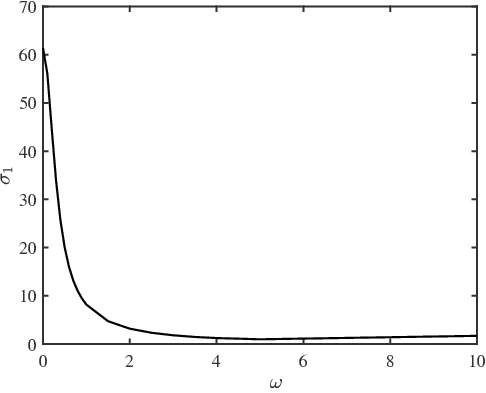}}\\
\vspace{0.2cm}
\subfloat[]{\includegraphics[width=0.45\textwidth]{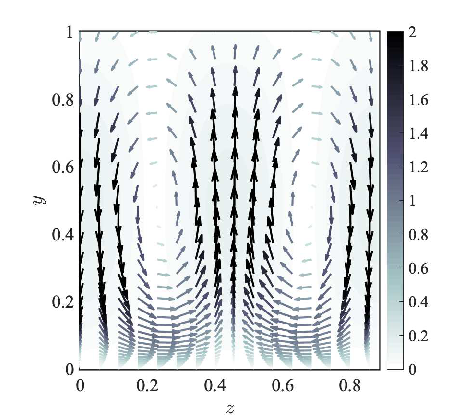}}
\hspace{0.3cm}
\subfloat[]{\includegraphics[width=0.45\textwidth]{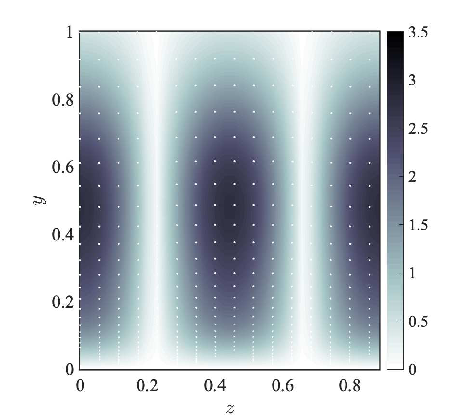}}
\caption{(a) Logarithm of spectral energy content, $\log(\hat{E})$, at
$y^+\approx 15$ for the buffer layer case. (b) Principal singular
value $\sigma_1$ as a function of $\omega$ for $(k_x^\circ,k_z^\circ)
= (0,1)$ for the buffer layer.  The $y$-$z$ plane of the principal (c)
forcing mode $\tilde{\boldsymbol{\phi}}_1$ and (d) response mode for
$(k_x^\circ,k_z^\circ,\omega) = (0,1,0)$ for the buffer layer case.
The streamwise component (colour) and the cross-flow component
(arrows) are given, with the colour bar indicating magnitude for both
components.
\label{fig:energy_dist}}
\end{center}
\end{figure}

As mentioned in the previous section, a choice of the target
wavenumbers is required to identify the forcing modes that are removed
at each time step. For this, we target the Fourier modes with the most
energy content. Figure \ref{fig:energy_dist}(a) shows the spectral
energy content, $\hat{E}(k_x,k_z) =
1/2\left(\hat{\boldsymbol{u}}^*\hat{\boldsymbol{u}}\right)$, at
$y^+\approx 15$ for the buffer layer case, as a function of streamwise
and spanwise wavenumber.  There is a clear peak at
$(k_x^\circ,k_z^\circ) = (0,\pm 1)$.  This is consistent with the fact
that the domain size of the minimal channel is such that it isolates
flow structures to be infinitely long in the streamwise direction and
once-periodic in the spanwise direction.  Thus, for our analysis, we
choose the streamwise and spanwise wavenumbers $(k_x^\circ,k_z^\circ)
= (0,1)$. This choice of streamwise and spanwise wavenumbers also
coincides with the $(k_x,k_z)$ with the largest $\sigma_1$ over all
values of $\omega$, which shows that not only does this wavenumber
pair hold the most energy, but also has the highest amplification
within the resolvent framework. The temporal frequency is given as
$\omega=0$, which corresponds to the highest $\sigma_1$ for
$(k_x^\circ,k_z^\circ) = (0,1)$ (figure \ref{fig:energy_dist}(b)).
Note that the projection \eqref{eq:projection} includes the
contributions from various temporal frequencies apart from $\omega =
0$ due to the fact that \eqref{eq:projection} is time-dependent.
However, it ascertains the removal of this particular forcing mode.
Also, the singular value associated with $\omega = 0$ is much larger
than other temporal frequencies, making the removal of other frequency
contents relatively less significant.

The principal forcing and response modes for this particular
frequency--wavenumber triplet are given in figure
\ref{fig:energy_dist}(c,d). The forcing mode highlights a pair of
streamwise rolls in the nonlinear term, i.e., the Reynolds stress
contribution to the advection term, spanning the entire channel
half-height. A much weaker streamwise streak in the nonlinear term
whose magnitude ($|\tilde{\boldsymbol{\phi}}_{1,u}|$) is approximately
5\% of that of the streamwise rolls,
$(\tilde{\boldsymbol{\phi}}_{1,v}^2+\tilde{\boldsymbol{\phi}}_{1,w}^2)^{1/2}$,
is also present. The response modes exhibit streamwise streaks with
alternating signs of the same magnitude (only the magnitude is shown
in figure \ref{fig:energy_dist}(d)) and weak cross flow. The values
above $y = 1$ are negligible due to our definition of the principal
mode. This shows that the streamwise rolls generated by the nonlinear
term can be transformed through a linear process, i.e., the lift-up
mechanism, to streamwise streaks, as expected from previous studies of
the SSP. For the given $(k_x,k_z,\omega)$, the principal forcing mode
contains the largest energetic contribution under unit broadband
forcing (approximately $85\%$), defined as
$(\sigma_i^2+\sigma_{i+1}^2)/\sum_{k=1}^\infty \sigma_k^2$ for each
$\tilde{\boldsymbol{\phi}_i}$ for $i=1,3,5,\cdots$. The subsequent
modes $\tilde{\boldsymbol{\phi}_3}$ and $\tilde{\boldsymbol{\phi}_5}$
have an energetic contribution of approximately $12\%$ and $2\%$,
respectively (see figure \ref{fig:TKE_180m_2}(a)). The large
separation in the singular values indicates that the principal forcing
mode will be amplified by almost an order of magnitude more than the
other forcing modes and thus will be integral in the SSP. 

\begin{figure}
\begin{center}
%\vspace{0.2cm}
\subfloat[]{\includegraphics[width=0.45\textwidth]{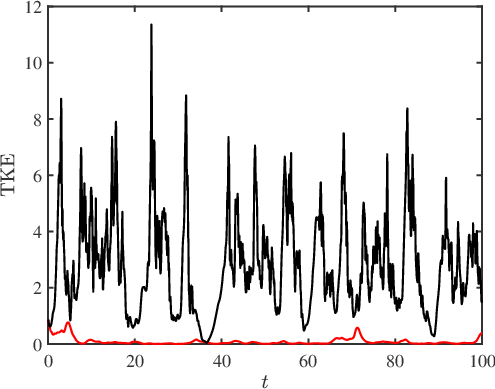}}
\hspace{0.3cm}
\subfloat[]{\includegraphics[width=0.45\textwidth]{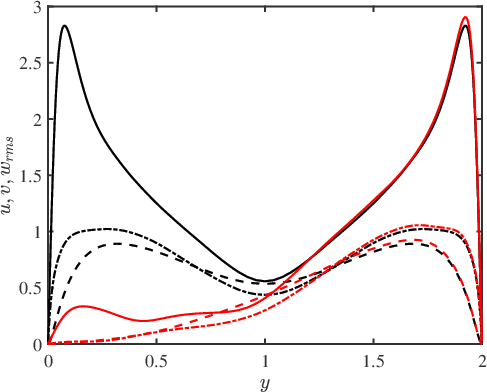}}\\
\subfloat[]{\includegraphics[width=0.45\textwidth]{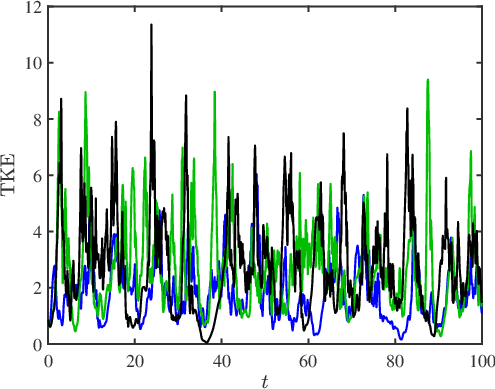}}
\hspace{0.3cm}
\subfloat[]{\includegraphics[width=0.45\textwidth]{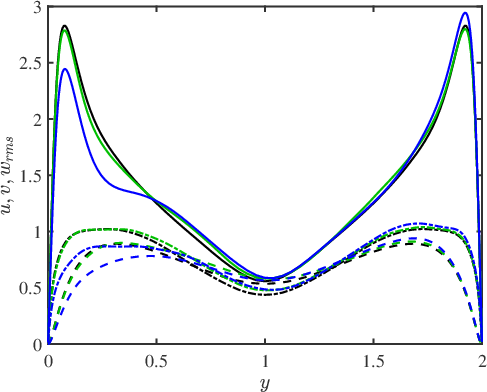}}
\caption{(a) Temporal evolution of TKE at $y^+\approx 15$ for the
damped buffer layer case removing $\hat{\boldsymbol{g}}_1^{(0,1,0)}$
(red) and the undamped buffer layer minimal
channel (black). (b) Streamwise (\solidline), wall-normal
(\dashline), and spanwise (\dotdashline) r.m.s. velocity fluctuations
for the damped case removing $\hat{\boldsymbol{g}}_1^{(0,1,0)}$ (red)
and the undamped (black) buffer layer minimal channel. (c,d) Same as
(a,b), but removing $\hat{\boldsymbol{g}}_3^{(0,1,0)}$
(blue) or $\hat{\boldsymbol{g}}_5^{(0,1,0)}$
(green) at each time step instead of
$\hat{\boldsymbol{g}}_1^{(0,1,0)}$.
\label{fig:TKE_180m}}
\end{center}
\end{figure}
The effect of removing $\hat{\boldsymbol{g}}_1^{(0,1,0)}$ can be seen
in figure \ref{fig:TKE_180m}(a) in the form of the turbulence kinetic
energy (TKE) evolution in time. Although not shown, the dissipation
rate as a function of time exhibits similar trends to that of the TKE.
Figure \ref{fig:TKE_180m}(a) shows that removing
$\hat{\boldsymbol{g}}_1^{(0,1,0)}$ reduces the TKE significantly.  The
steady-state root-mean-square (r.m.s.) velocity fluctuation profiles
for this case are given in figure \ref{fig:TKE_180m}(b). As expected,
the effect of removing the principal forcing term is observed only on
the bottom half of the channel where the principal forcing term was
isolated, with only minor changes in the statistics in the top half of
the channel. At any instance in time of the undamped case, the
contribution of $\hat{\boldsymbol{g}}_1^{(0,1,0)}$ to the nonlinear
advection term, defined as the average-in-time ratio of $\int
\hat{\boldsymbol{g}}_1^{(0,1,0)*}\hat{\boldsymbol{g}}_1^{(0,1,0)}
\mathrm{d}y$ to $\iiint \boldsymbol{f}'^2
\mathrm{d}x\mathrm{d}y\mathrm{d}z$, is less than $0.9\%$, and removing
the same magnitude randomly from the advection term at each time step
had no effect on the one-point statistics, which support the
importance of the spatial structure of the mode being projected out to
the turbulent flow. Furthermore, if we remove
$1/2\hat{\boldsymbol{g}}_1^{(0,1,0)}$ at each time step rather than
$\hat{\boldsymbol{g}}_1^{(0,1,0)}$, the impact on the turbulence
statistics is negligible compared to the undamped case, which further
highlights the importance of the principal forcing term.

\begin{figure}
\begin{center}
%\vspace{0.2cm}
\subfloat[]{\includegraphics[width=0.45\textwidth]{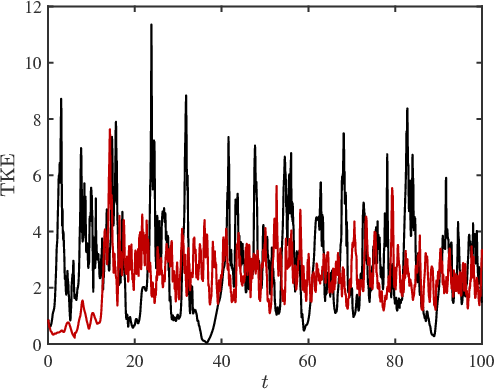}}
\hspace{0.3cm}
\subfloat[]{\includegraphics[width=0.45\textwidth]{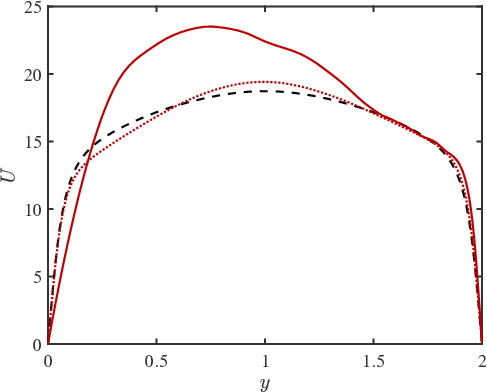}}
\caption{(a) Temporal evolution of TKE at $y^+\approx 15$ for the
damped buffer layer case removing $\hat{\boldsymbol{g}}_1^{(0,1,0)}$
without fixing the mean (maroon) and the undamped
buffer layer minimal channel (black). (b) Mean streamwise
velocity corresponding to $t = 12$ (maroon;\solidline) 
and averaged over $20<t<100$ (maroon;\dotline) for the
case without fixing the mean and the fixed mean profile 
(black;\dashline). 
\label{fig:TKE_180m_Uch}} 
\end{center}
\end{figure}
In the case where the mean flow is not frozen but rather allowed to
evolve in time (figure \ref{fig:TKE_180m_Uch}(a)), the time evolution
of TKE follow similar results as shown in figure \ref{fig:TKE_180m}(a)
until $t\approx 12$, where it starts to increase again.  This is due
to the change in the mean velocity profile, as shown in figure
\ref{fig:TKE_180m_Uch}(b). The initial reduction of turbulence
intensities changes the dynamics of the flow such that the the mean
velocity profile approaches the laminar profile on the bottom half of
the channel. The drastic change in the mean profile invalidates the
relevance of the the principal forcing mode obtained from the
turbulent mean state, $U^F$, and causes the TKE to increase again.
Even after the TKE becomes statistically stationary, $t>20$, the new
equilibrium state after transients shown in \ref{fig:TKE_180m_Uch}(b),
differs from the turbulence with mean profile $U^F$.

In order to study the effects of removing subsequent forcing modes, we
repeat the previous frozen-mean experiment, but removing either
$\hat{\boldsymbol{g}}_3^{(0,1,0)}$ or
$\hat{\boldsymbol{g}}_5^{(0,1,0)}$, the secondary and tertiary forcing
modes restricted primarily to the bottom half of the channel as
defined in \S\ref{sec:methods:resolvent}, instead of
$\hat{\boldsymbol{g}}_1^{(0,1,0)}$. We see that turbulence is
sustained in both cases from figure \ref{fig:TKE_180m}(c), but the
extreme peaks in TKE observed in the undamped case are not as
prominent. We can also see in figure \ref{fig:TKE_180m}(d) that while
the effect of removing $\hat{\boldsymbol{g}}_3^{(0,1,0)}$ still has
some impact on the steady-state turbulence intensities, especially
around $y^+\approx 15$, the net change in the statistics is much
smaller than that of removing $\hat{\boldsymbol{g}}_1^{(0,1,0)}$.
Removing $\hat{\boldsymbol{g}}_5^{(0,1,0)}$ has no impact on one-point
statistics, and similar results are expected of subsequent forcing
modes. Additionally, removing
$\hat{\boldsymbol{g}}_3^{(0,1,0)}$ or
$\hat{\boldsymbol{g}}_5^{(0,1,0)}$ in conjunction with
$\hat{\boldsymbol{g}}_1^{(0,1,0)}$ (not shown) resulted in similar
statistics as removing only $\hat{\boldsymbol{g}}_1^{(0,1,0)}$. At
any instance in time, the average contribution of
$\hat{\boldsymbol{g}}_3^{(0,1,0)}$ or
$\hat{\boldsymbol{g}}_5^{(0,1,0)}$ in the undamped case are
statistically similar to the contribution of
$\hat{\boldsymbol{g}}_1^{(0,1,0)}$ at $0.9\%$ of the total advection
term, which shows the dominant impact of the principal forcing mode on
the SSP.  
\begin{figure}
\begin{center}
%\vspace{0.2cm}
\subfloat[]{\includegraphics[width=0.45\textwidth]{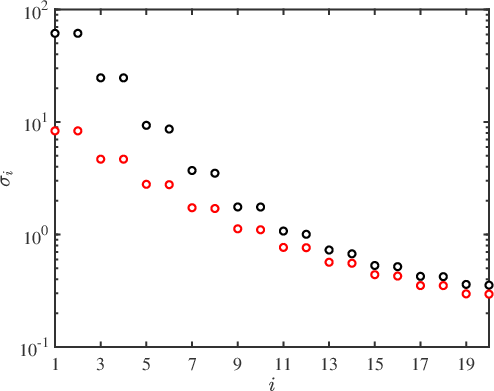}}
\hspace{0.3cm}
\subfloat[]{\includegraphics[width=0.45\textwidth]{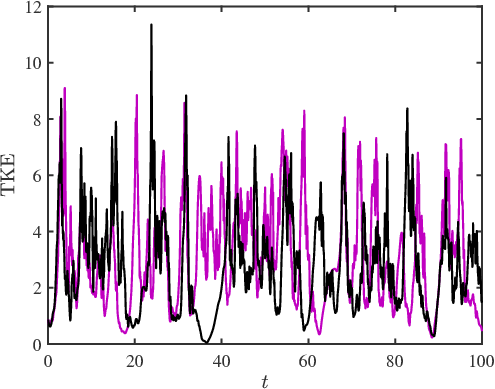}}\\
\caption{(a) Amplification factors $\sigma_i$ for
$(k_x^\circ,k_z^\circ,\omega)=(0,1,0)$ (black) and
$(k_x^\circ,k_z^\circ,\omega)=(0,2,0)$ (red) for the buffer layer
case. (b) Temporal evolution of TKE at $y^+\approx 15$ for the damped
buffer layer case removing $\hat{\boldsymbol{g}}_1^{(0,2,0)}$
(magenta) and the undamped buffer layer minimal
channel (black). 
\label{fig:TKE_180m_2}}
\end{center}
\end{figure}
Additionally, we also repeat the experiment, removing
$\hat{\boldsymbol{g}}_1^{(0,2,0)}$, as $(k_x^\circ,k_z^\circ)=(0,2)$
contains a significant amount of energy in figure
\ref{fig:energy_dist}(a). However, compared to the triplet
$(k_x^\circ,k_z^\circ,\omega)=(0,1,0)$, the amplification factor
$\sigma_1$ for $(k_x^\circ,k_z^\circ,\omega)=(0,2,0)$ is an order of
magnitude lower as seen in figure \ref{fig:TKE_180m_2}(a) and the
principal mode has an energetic contribution of approximately 65\%,
which is much lower than the 85\% for
$(k_x^\circ,k_z^\circ,\omega)=(0,1,0)$. The resulting time evolution
of TKE shown in figure \ref{fig:TKE_180m_2}(b) is hence much less
affected than removing $\hat{\boldsymbol{g}}_1^{(0,1,0)}$
and removing $\hat{\boldsymbol{g}}_1^{(0,2,0)}$ in
conjunction with $\hat{\boldsymbol{g}}_1^{(0,1,0)}$ (not shown)
resulted in marginal difference compared to removing only
$\hat{\boldsymbol{g}}_1^{(0,1,0)}$. Overall, the removal of the
principal forcing mode corresponding to
$(k_x^\circ,k_z^\circ,\omega)=(0,1,0)$ has the most impact on the
flow, as expected.   

%----------------------------------------------------------------%
\subsection{Logarithmic layer}\label{sec:results:2000l}
%----------------------------------------------------------------%

\begin{figure}
\begin{center}
\vspace{0.12cm}
\subfloat[]{\includegraphics[width=0.45\textwidth]{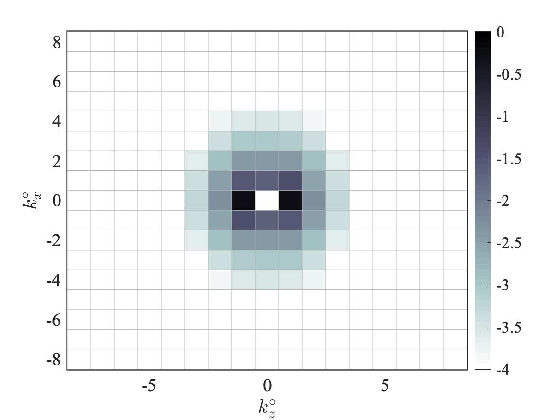}}
\hspace{0.3cm}
\subfloat[]{\includegraphics[width=0.40\textwidth]{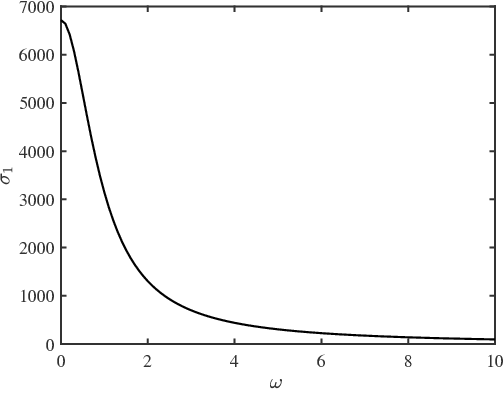}}\\
\vspace{0.2cm}
\subfloat[]{\includegraphics[width=0.45\textwidth]{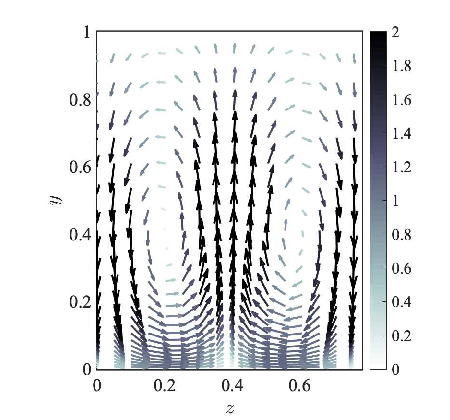}}
\hspace{0.3cm}
\subfloat[]{\includegraphics[width=0.45\textwidth]{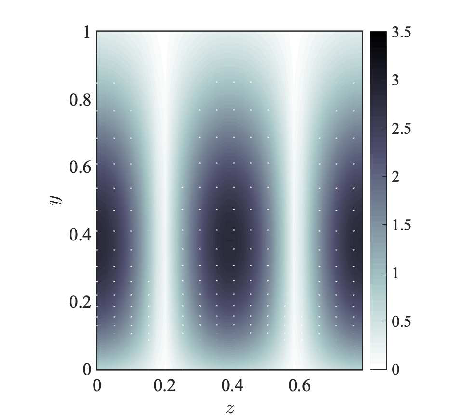}}
\caption{(a) Logarithm of spectral energy content, $\log(\hat{E})$, at
$y\approx 0.15$ for the log-layer case. (b) Principal singular
value $\sigma_1$ as a function of $\omega$ for $(k_x^\circ,k_z^\circ)
= (0,1)$ for the log-layer case.  The $y$-$z$ plane of the principal
(c) forcing mode $\tilde{\boldsymbol{\phi}}_1$ and (d) response mode
for $(k_x^\circ,k_z^\circ,\omega) = (0,1,0)$ for the log-layer case.
The streamwise component (colour) and the cross-flow component
(arrows) are given, with the colour bar indicating magnitude for both
components. 
\label{fig:energy_dist_log}}
\end{center}
\end{figure}

Using the same approach as the buffer layer case, we identify the
target wavenumbers for the log layer case. Figure
\ref{fig:energy_dist_log}(a) shows $\hat{E}(k_x,k_z) =
1/2\left(\hat{\boldsymbol{u}}^*\hat{\boldsymbol{u}}\right)$ at $y
\approx 0.15$ for the log layer case, as a function of streamwise and
spanwise wavenumber. Again, there is a clear peak at
$(k_x^\circ,k_z^\circ) = (0,\pm 1)$, which is consistent with the fact
that the domain size of the minimal channel for the log layer is such
that the most dominant flow structures are infinitely long in the
streamwise direction and once-periodic in the spanwise direction.
Thus, for our analysis, we again choose the streamwise and spanwise
wavenumbers $(k_x^\circ,k_z^\circ) = (0,1)$.  However, note that
unlike the buffer layer case, the spectral energy content of the
log-layer case is distributed among various combinations of $k_x$ and
$k_z$, in particular with large contributions from
$(k_x^\circ,k_z^\circ)=(\pm1,\pm1)$, indicating a more complex
phenomenon in the SSP that cannot be isolated to one wavenumber, as
seen in figure \ref{fig:energy_dist_log}(a).  However, similar to the
analysis on $(k_x^\circ,k_z^\circ) = (0,2)$ for the buffer layer case,
the largest singular values as well as the spectral gap for the
non-fundamental wavenumbers are significantly lower than that of
$(k_x^\circ,k_z^\circ)=(0,1)$, indicating that the effect of the
principal resolvent modes for $(k_x^\circ,k_z^\circ)\neq(0,1)$ are not
as dominant as that of $(k_x^\circ,k_z^\circ)=(0,1)$.  The temporal
frequency identified by the highest $\sigma_1$ for
$(k_x^\circ,k_z^\circ) = (0,1)$ (figure \ref{fig:energy_dist_log}(b))
is $\omega = 0$, but the values of $\sigma_1$ do not decay as fast as
the buffer layer case, indicating that more combinations of $\omega$
may be involved compared to the buffer layer case.

The principal forcing and response modes for this particular
frequency--wavenumber triplet are given in figure
\ref{fig:energy_dist_log}(c,d). Despite the different Reynolds number
and the wall boundary condition, the forcing and response modes of the
log layer are similar to those of the buffer layer. The forcing mode
highlights a pair of streamwise rolls spanning the entire channel
half-height, and the response modes exhibit streamwise streaks with
alternating signs of the same magnitude. For the given
$(k_x,k_z,\omega)$, the principal forcing mode contains the largest
energetic contribution under unit broadband forcing (approximately
$81\%$), similar to the buffer layer case. The similarity in the
identified principal resolvent modes indicate the similarity in the
underlying mechanism for the SSP of the buffer and log layers, as it
has been hypothesised in the literature.

The fact that the log layer structure of the minimal flow
unit spans the full boundary layer thickness can be further explained
by the self-similar attached eddy hypothesis
\citep{Townsend1976,Perry1982,Meneveau2013,Agostini2017,Marusic2019}.
\citet{McKeon2019} has recently identified that an isolated resolvent
mode in with a given convection velocity constitutes only part of the
full attached eddy, and a geometric progression of such modes needs to
be considered for the fair comparison to the attached eddies.  In
particular, a single wall-normal location of the log layer will be
affected by a series of wall-attached eddies whose footprint includes
the given wall-normal height. From this viewpoint, the structures that
impact the log layer would span the whole half channel as observed in
the current case.

\begin{figure}
\begin{center}
%\vspace{0.2cm}
\subfloat[]{\includegraphics[width=0.45\textwidth]{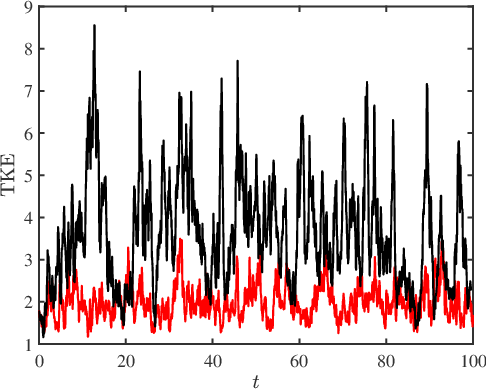}}
\hspace{0.3cm}
\subfloat[]{\includegraphics[width=0.45\textwidth]{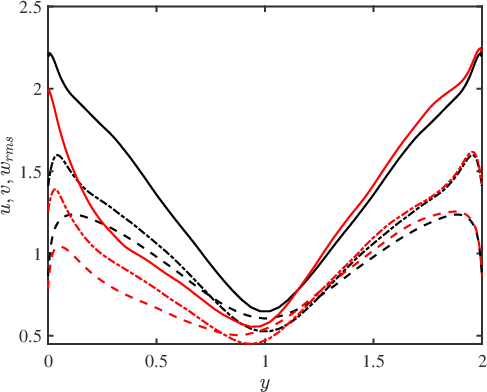}}\\
\caption{(a) Temporal evolution of TKE at $y\approx 0.15$ for the
damped log layer case removing $\hat{\boldsymbol{g}}_1^{(0,1,0)}$
(red) and the undamped log layer minimal channel
(black). (b) Streamwise (\solidline), wall-normal (\dashline),
and spanwise (\dotdashline) r.m.s. velocity fluctuations for the damped
log layer case removing $\hat{\boldsymbol{g}}_1^{(0,1,0)}$
(red) and the undamped (black) log layer minimal channel. 
\label{fig:TKE_2000l}}
\end{center}
\end{figure}

The results of removing $\hat{\boldsymbol{g}}_1^{(0,1,0)}$ for the log
layer are shown in figure \ref{fig:TKE_2000l}. The TKE does not reduce
as significantly as the buffer layer case, as seen in the time
evolution of TKE in figure \ref{fig:TKE_2000l}(a) and the
time-averaged turbulence intensities in figure \ref{fig:TKE_2000l}(b);
however, we do see a significant reduction in the TKE. 
Similar to the buffer layer case, the dissipation rate
(not shown) follow the trends of the TKE. The relatively smaller
effect of the principal forcing modes in the log layer is due to the
complexity and the disorganisation of the log layer itself.  Unlike
the buffer layer, where only one scale is isolated by the minimal
domain, the log layer still includes contributions from various scales
that exist in the log layer (figure \ref{fig:energy_dist_log}(a)),
which makes it harder to affect the full system by removing only one
scale. However, the significant reduction in TKE and turbulence
intensities obtained by removing the principal forcing mode, which
exhibit similar characteristics in both the buffer and log layer,
shows that the mechanism in which turbulence self-sustains is similar
in both regions. The results of removing secondary
forcing modes and principal forcing modes for a different
$(k_x,k_z,\omega)$, while not reported here, show comparable trends as
seen in the buffer layer, where the impact on the flow statistics are
marginal compared to that of $\hat{\boldsymbol{g}}_1^{(0,1,0)}$.

%----------------------------------------------------------------%
\section{Nonlinear interaction}\label{sec:nonlinear}
%----------------------------------------------------------------%

\begin{figure}
\begin{center}
%\vspace{0.2cm}
\includegraphics[width=0.45\textwidth]{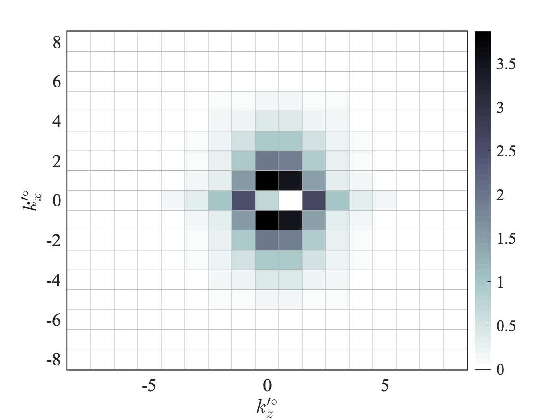}
\includegraphics[width=0.45\textwidth]{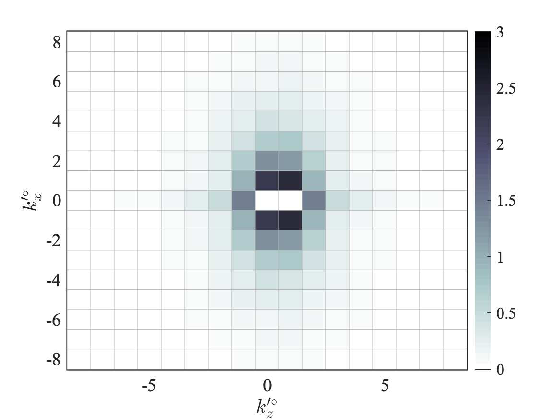}
\caption{Average contribution of each convolution sum
$|\Pi(k'_x=1,k'_z=0;k_x=0,k_z=1,\omega=0)|$ normalised by the total
contribution
$\left|\sum_{k'_x,k'_z}\Pi(k'_x,k'_z;k_x=0,k_z=1,\omega=0)\right|$ for
the (a) buffer layer case and (b) log layer case.
\label{fig:fourier_contribution}}
\end{center}
\end{figure}

As demonstrated in the previous section, the principal forcing mode
$\tilde{\boldsymbol{\phi}}_1^{(0,1,0)}$ identifies the most amplified
nonlinear interaction for the most energetic wavenumber and is
integral in sustaining turbulence in the near wall cycle for both the
buffer and log layers. In order to study the nonlinear interactions
that produce this term through dyadic interactions, we decompose the
nonlinear term as a convolution sum in Fourier space
\begin{equation}
\hat{\boldsymbol{f}}(k_x,k_z) =
\sum_{k'_x,k'_z=-\infty}^\infty
\left(\hat{\boldsymbol{u}}(k'_x,k'_z)\cdot\hat{\nabla}\right)
\hat{\boldsymbol{u}}(k_x-k'_x,k_z-k'_z),
\end{equation}
which is a function of $y$ and $t$.  The contribution of each
component of the convolution sum toward the projection of the
principal forcing term onto $\tilde{\boldsymbol{\phi}}_1^{(0,1,0)}$
for any given flow field can be measured as
\begin{equation}
\Pi(k'_x,k'_z;k_x,k_z,\omega) = \left\langle
\left[\begin{array}{c}
\left(\hat{\boldsymbol{u}}(k'_x,k'_z)\cdot\hat{\nabla}\right)
\hat{\boldsymbol{u}}(k_x-k'_x,k_z-k'_z)\\ 
0
\end{array}\right],
\tilde{\boldsymbol{\phi}}_1(k_x,k_z,\omega)\right\rangle.
\end{equation}
Note that integration over all $k'_x$ and $k'_z$ of
$\Pi(k'_x,k'_z)^{(0,1,0)}$ gives the projection coefficient computed
in Eq. \eqref{eq:projection}, where by abuse of notation, we define
$\Pi(k'_x,k'_z)^{(a,b,c)}=\Pi(k'_x,k'_z;k_x^\circ=a,k_z^\circ=b,\omega=c)$
and
$\Pi^{(a,b;c,d,e)}=\Pi(k^{\prime\circ}_x=a,k^{\prime\circ}_z=b;k_x^\circ=c,k_z^\circ=d,\omega=e)$.
Also, due to incompressibility, $\Pi$ is symmetric with respect to
$(k_x,k_z)$ for each flow field; that is,
$\Pi(k'_x,k'_z;k_x,k_z,\omega) =
\Pi(k_x-k'_x,k_z-k'_z;k_x,k_z,\omega)$.  The average spectral map of
$|\Pi(k'_x,k'_z)^{(0,1,0)}|$ normalised by the total contribution
$|\sum_{k_x',k_z'}\Pi(k'_x,k'_z)^{(0,1,0)}|$ is computed from flow
fields of the undamped minimal channel of the buffer and log layer and
is depicted in figure \ref{fig:fourier_contribution}.  Both spectral
maps show quantitatively similar features and identify two main
sources of contribution from the wavenumber pair
$(k^{\prime\circ}_x,k^{\prime\circ}_z) = (1,0)$ and
$((k_x-k'_x)^\circ,(k_z-k'_z)^\circ) = (-1,1)$ and its mirror image in
the $x$-plane, $(k^{\prime\circ}_x,k^{\prime\circ}_z) = (-1,1)$ and
$(-k^{\prime\circ}_x,1-k^{\prime\circ}_z) = (1,0)$, which account for
approximately $40\%$ of the total contribution. This is consistent
with the results from \citet{Hamilton1995}, which used vorticity as a
measure rather than the contribution towards resolvent forcing modes.
While the contributions from other wavenumber pairs are not
negligible, for the remainder of this paper, we focus on the two pairs
of wavenumbers to identify the coherent structures responsible for the
nonlinear forcing term and without loss of generality choose the pair
$(k^{\prime\circ}_x,k^{\prime\circ}_z) = (1,0)$ and
$((k_x-k'_x)^\circ,(k_z-k'_z)^\circ) = (-1,1)$.  

To identify instantaneous flow configurations where the contribution
towards $\tilde{\boldsymbol{\phi}}_1^{(0,1,0)}$ is strong or weak, we
first observe values of $\bar\Pi^{(1,0;0,1,0)} =
|\Pi^{(1,0;0,1,0)}|/\langle\hat{\boldsymbol{f}}(k^{\prime\circ}_x=1,k^{\prime\circ}_z=0),
\hat{\boldsymbol{f}}(k^{\prime\circ}_x=1,k^{\prime\circ}_z=0)\rangle^{1/2}$,
which represents the normalised contribution to the principal forcing
term with respect to the total energy of the nonlinear term in the
$(k'_x,k'_z)$  mode, computed from flow fields of the undamped
channel. We then determine the mean $\mu$ and standard deviation
$\varsigma$ of the distribution of $\bar\Pi^{(1,0;0,1,0)}$ over all
time instances. The high forcing-intensity events are defined as those
with $\bar\Pi^{(1,0;0,1,0)} > \mu+2\varsigma$ and low
forcing-intensity events as those with $\bar\Pi^{(1,0;0,1,0)} <
\mu-2\varsigma$. Both cases consist of approximately 5\% of the total
events. 

\begin{figure}
\begin{center}
%\vspace{0.2cm}
{\includegraphics[height=0.26\textwidth]{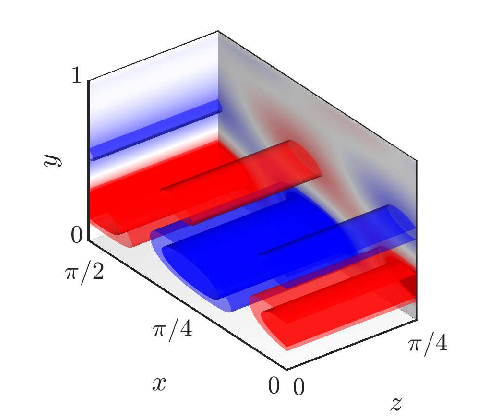}}
{\includegraphics[height=0.26\textwidth]{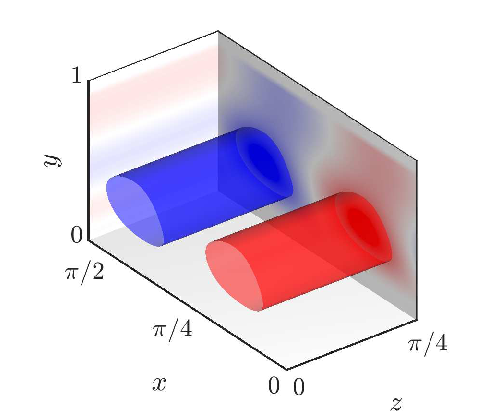}}
{\includegraphics[height=0.26\textwidth]{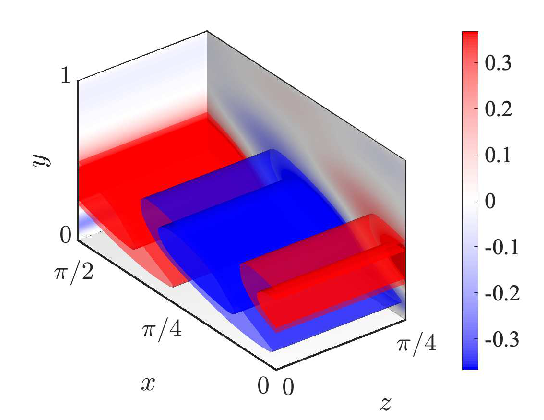}}
\\
{\includegraphics[height=0.26\textwidth]{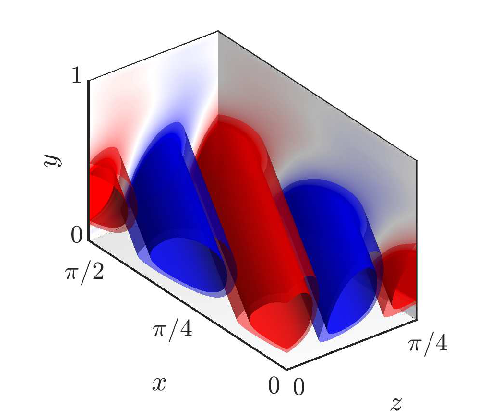}}
{\includegraphics[height=0.26\textwidth]{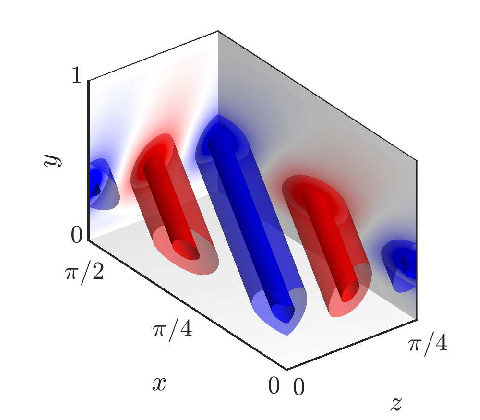}}
{\includegraphics[height=0.26\textwidth]{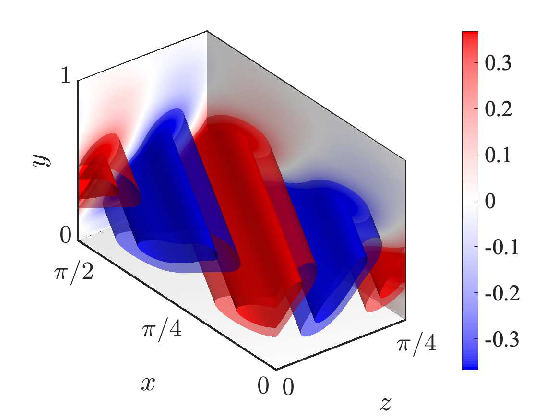}}
\caption{Average $\mathcal{F}^{-1}\left(\hat{u}\right)$ (left),
$\mathcal{F}^{-1}\left(\hat{v}\right)$ (centre),
$\mathcal{F}^{-1}\left(\hat{w}\right)$ (right) for
$(k^\circ_x,k^\circ_z)=(1,0)$ (top) and $(-1,1)$ (bottom) conditioned to high
forcing-intensity events for the buffer layer case. The isosurfaces
are $0.41$ (solid red), $0.26$ (transparent red), $-0.26$ (transparent
blue), and $-0.41$ (solid blue).
\label{fig:strong_modes}}
\end{center}
\end{figure}

\begin{figure}
\begin{center}
%\vspace{0.2cm}
{\includegraphics[height=0.24\textwidth]{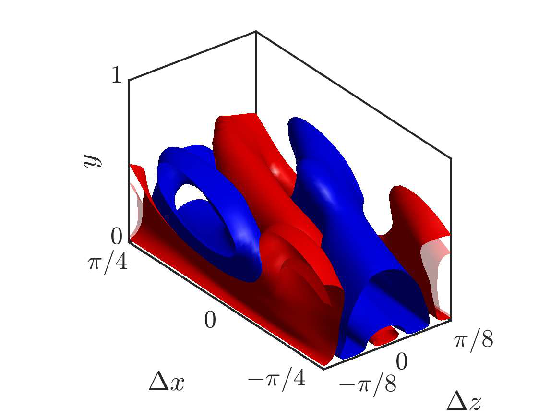}}
{\includegraphics[height=0.24\textwidth]{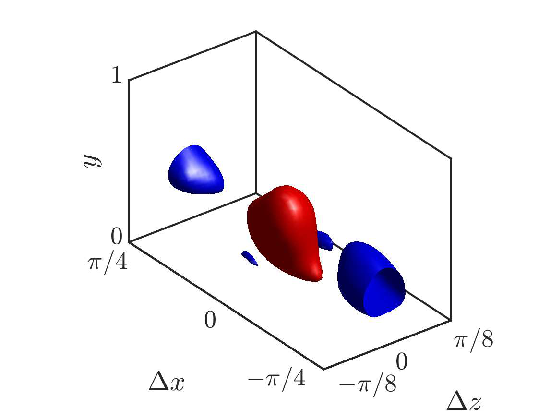}}
{\includegraphics[height=0.24\textwidth]{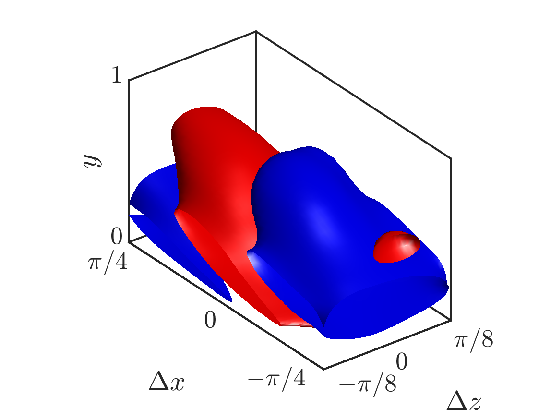}}\\
%\vspace{0.2cm}
{\includegraphics[height=0.24\textwidth]{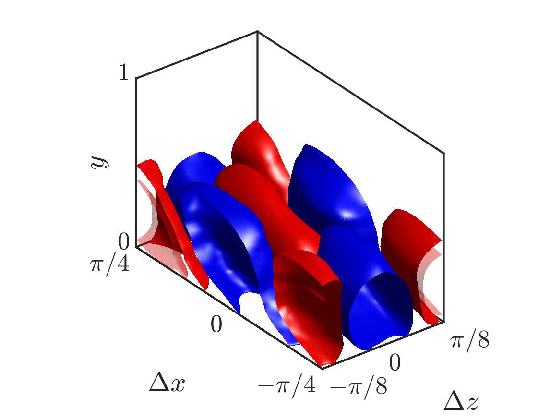}}
{\includegraphics[height=0.24\textwidth]{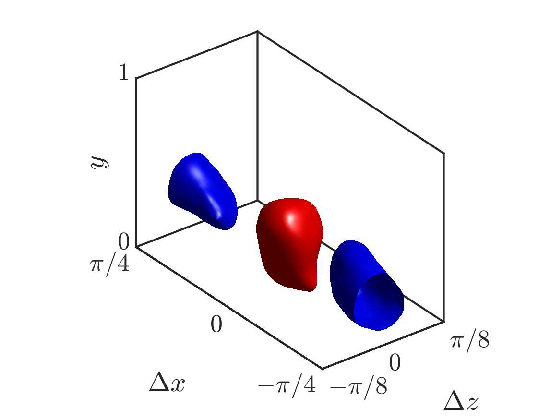}}
{\includegraphics[height=0.24\textwidth]{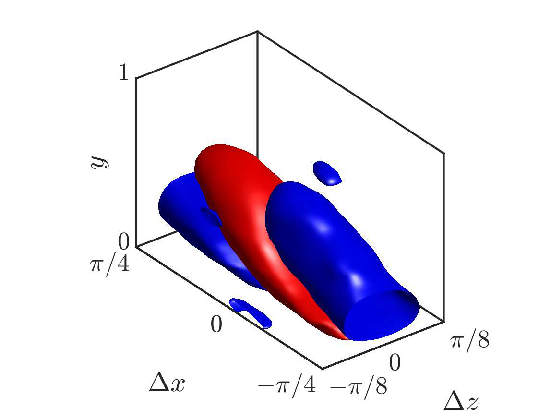}}
\caption{Correlations $C_{uu}$ (left), $C_{vv}$ (centre), and $C_{ww}$
(right) conditioned to high forcing-intensity events for the buffer
(top) and log layer (bottom).  The isosurfaces are $0.1$ (red) and
$-0.04$ (blue).
\label{fig:strong_correlations}}
\end{center}
\end{figure}

\begin{figure}
\begin{center}
%\vspace{0.2cm}
{\includegraphics[height=0.24\textwidth]{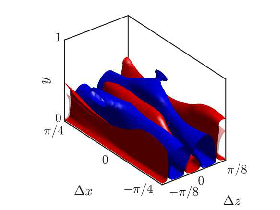}}
{\includegraphics[height=0.24\textwidth]{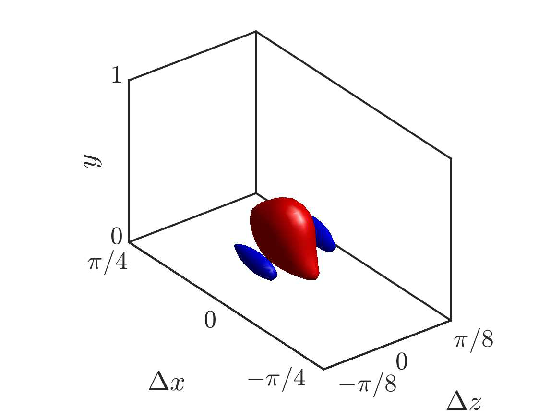}}
{\includegraphics[height=0.24\textwidth]{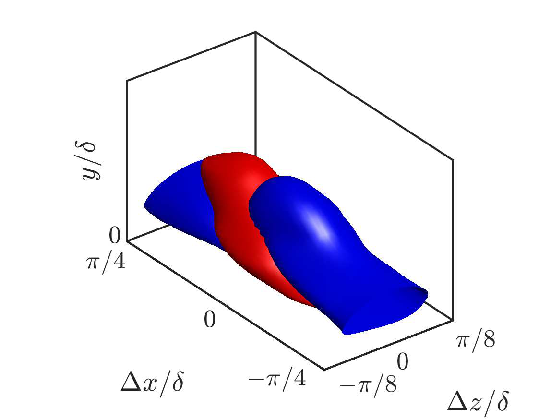}}\\
%\vspace{0.2cm}
{\includegraphics[height=0.24\textwidth]{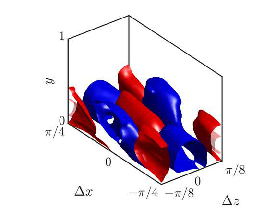}}
{\includegraphics[height=0.24\textwidth]{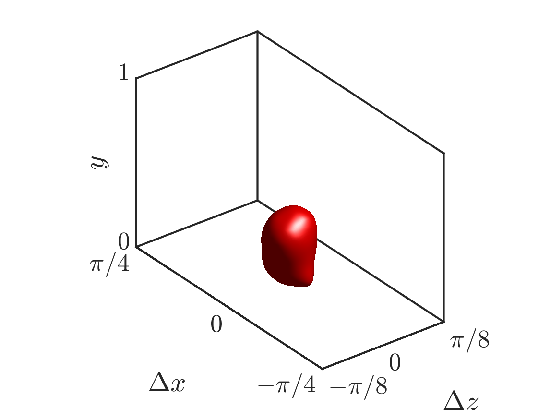}}
{\includegraphics[height=0.24\textwidth]{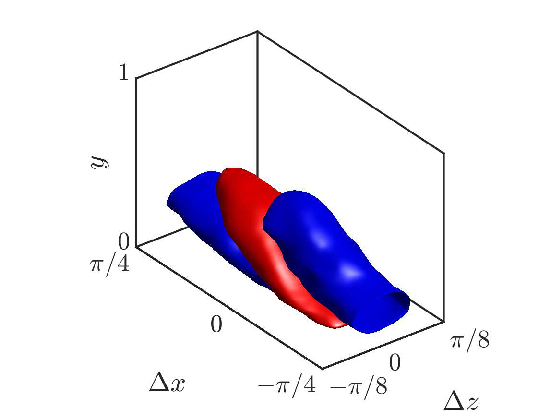}}
\caption{Correlations $C_{uu}$ (left), $C_{vv}$ (centre), and $C_{ww}$
(right) conditioned to low forcing-intensity events for the buffer
(top) and log layer (bottom).  The isosurfaces are $0.1$ (red) and
$-0.04$ (blue).
\label{fig:weak_correlations}}
\end{center}
\end{figure}

The average
$\mathcal{F}^{-1}\left(\hat{\boldsymbol{u}}(k^\circ_x=1,k^\circ_z=0)\right)$
and
$\mathcal{F}^{-1}\left(\hat{\boldsymbol{u}}(k^\circ_x=-1,k^\circ_z=1)\right)$
conditioned to high forcing-intensity events for the buffer layer are
shown in figure \ref{fig:strong_modes}, where $\mathcal{F}^{-1}$ is
the inverse Fourier transform. The modes are phase shifted before
averaging such that they are phase aligned for the streamwise velocity
component at $y^+\approx40$ in an effort to centre the underlying
structures at a fixed location. The coherent structures identified by
the $(k_x^\circ,k_z^\circ) = (1,0)$ mode (top row of figure 9) are in
the form of a pair of spanwise rolls that is being sheared in the
spanwise direction by $w$. That is, the relative phase
of the velocity components depict the vertically displaced pairs of
positive and negative streamwise velocity components located between
the horizontally displaced pairs of positive and negative wall-normal
velocity form structure in the shape of a spanwise roll, which then
are affected by the alternating sign of the $w$ components. The
$(k_x^\circ, k_z^\circ) = (-1, 1)$ mode (bottom row of figure 9) show
oblique streaks, which are the components of meandering streaks, with
high-speed streaks moving towards the wall and low-speed streaks
moving away from the wall, corresponding to sweeps and ejections.
Furthermore, the spatial auto-correlation coefficients of the
streamwise, wall-normal, and spanwise velocity fluctuations, denoted
$C_{uu}$, $C_{vv}$, and $C_{ww}$ respectively, are computed, where
\begin{equation}
C_{qq}(x-x',y,z-z') = \frac
{\mathbb{E}[q(x,y,z)q(x',y',z')|\bar\Pi^{(1,0;0,1,0)} > \mu+2\varsigma]}
{\mathbb{E}[q(x,y,z)q(x,y,z)|\bar\Pi^{(1,0;0,1,0)} > \mu+2\varsigma]},
\end{equation} 
is the correlation conditioned to high forcing-intensity events at
$y'^+\approx 40$ for the buffer layer and $y' \approx 0.2$ for the log
layer.  Here , $q\in\{u,v,w\}$ and $\mathbb{E}[\cdot]$ is the
expectation, or alternatively the average quantity over homogeneous
spatial directions and instances. The correlations for both cases,
shown in figure \ref{fig:strong_correlations}, reveal structures very
similar to ones highlighted by the time-averaged
$\mathcal{F}^{-1}\left(\hat{\boldsymbol{u}}(k^\circ_x=1,k^\circ_z=0)\right)$
and
$\mathcal{F}^{-1}\left(\hat{\boldsymbol{u}}(k^\circ_x=-1,k^\circ_z=1)\right)$
and resemble spanwise-sheared spanwise rolls ($C_{vv}$ and $C_{ww}$)
with oblique streaks ($C_{uu}$), which are shared for both the buffer
and log layer. This indicates that not only the linear mechanism but
also the nonlinear mechanism driving the SSP are similar in both
regions with similar structures forming the precursor events leading
to strong intensities of the principal resolvent forcing term.

On the contrary, although not shown, average
$\mathcal{F}^{-1}\left(\hat{\boldsymbol{u}}(k^\circ_x=1,k^\circ_z=0)\right)$
and
$\mathcal{F}^{-1}\left(\hat{\boldsymbol{u}}(k^\circ_x=-1,k^\circ_z=1)\right)$
conditioned to low forcing-intensity events are less coherent for
these wave parameters. This can be observed from the conditional
correlation of the buffer layer case in the top row of figure
\ref{fig:weak_correlations}, where the streamwise and wall-normal
velocities only show structures resembling straight streaks
corresponding to wavenumbers $(k^\circ_x,k^\circ_z) = (0,2)$ and no
structures resembling oblique streaks or spanwise rolls.  Considering
the fact that the total kinetic energy distribution for the strong and
weak events are similar, these results show that the precursor to the
nonlinear interaction that generates the principal forcing mode has
more defined coherent structures in the form of spanwise rolls and
oblique streaks, which interact to produce the principal forcing mode,
which then plays an important role in the SSP of near-wall turbulence.
Identifying these precursor events will allow development of new
control mechanisms that aim to reduce the production of the resolvent
forcing modes essential in sustaining turbulence.

Unlike the buffer layer case, the streamwise correlation conditioned
to low forcing-intensity events shown in the bottom row of figure
\ref{fig:weak_correlations} is not as different from the high
forcing-intensity events for the log layer case. The wall-normal
correlation still shows a clear positive and negative correlation
pairs in the streamwise direction when conditioned to low
forcing-intensity events, and the spanwise correlation are less
pronounced in the spanwise direction. The similarity in the streamwise
component, although less pronounced in the low forcing-intensity
events, shows that even if the principal forcing mode is removed, the
flow will still self-sustain structures necessary for regenerating
streamwise vortices, which is corroborated in
\S\ref{sec:results:2000l}. However, the most significant difference in
the correlations conditioned to strong and weak forcing-intensity
events are the wall-normal component that partly form the spanwise
rolls. These spanwise rolls are observed in cases of drag increase in
active and passive control
\citep{Garcia-Mayoral2011,Garcia-Mayoral2012,Toedtli2019}, which is
consistent with our observation that these are precursor events to
generation of the principal resolvent forcing term that is crucial in
generating and self-sustaining turbulence.

%----------------------------------------------------------------%
\section{Conclusions}\label{sec:conclusions}
%----------------------------------------------------------------%

We have studied the SSP of wall-bounded turbulence in the buffer and
log layers with special emphasis on the nonlinear mechanisms involved
in vortex regeneration. For this purpose, we have utilised resolvent
analysis to identify the most amplified nonlinear term in the
incompressible Navier-Stokes equations and studied the effect of this
term on numerical simulations of turbulent channel flow tailored to 
study isolated structures in the buffer and log layer.

Simulations of the minimal channel for the buffer and log layer with a
fixed mean streamwise velocity profile were performed to isolate the
structures at a prescribed scale. The most amplified nonlinear term
corresponding to the most energetic wavenumber was then computed from
the resolvent analysis using the mean velocity profile of the minimal
channel simulations. The identified mode was removed from the
nonlinear term of the corresponding simulation for a minimal channel
simulation with a fixed mean velocity profile at each time step. This
is made possible by the orthonormality of the resolvent modes, which
provides an orthonormal basis for the nonlinear term to be projected
onto. We have shown that the removal of the principal forcing mode
leads to a reduction of turbulence in the flow. We also applied the
removal method for subsequent forcing modes as well as non-fundamental
wavenumbers instead and observed only a marginal decrease in the
turbulence intensities, which reinstates the principal forcing mode
for the fundamental wavenumber as the most amplified, and thus the
most important, component of the nonlinear term. The diminished effect
on the log layer is attributed to the multi-scale nature of
high-Reynolds-number turbulence due to the more complicated
interaction between various scales.

Finally, we identified the coherent structures that, through the
nonlinear interaction, form the principal forcing mode. The identified
structures are in the form of spanwise-sheared spanwise rolls and
oblique streaks for both the buffer and log layer. The interaction of
the two components highlighted here regenerates streamwise vortices,
which through the lift-up mechanism amplifies streamwise streaks.
These streamwise streaks break down, spawning new generations of
meandering streaks and spanwise rolls, completing the SSP. The
similarities in the structures identified for the buffer and log layer
highlights the similarity of the nonlinear process in the SSP of the
two cases.

This new technique allows the analysis of nonlinear mechanisms 
by combining resolvent analysis with numerical simulations of 
fully non-linear Navier—Stokes equations in a realistic turbulent 
flow. The results show that the nonlinear mechanism of the buffer and 
log layer are
driven by similar structures and that without these structures present
in the flow, turbulence intensity is significantly reduced.  The
findings corroborate previous studies on nonlinear interaction of the
SSP and allow the characterisation of the underlying quadratic
interactions in the SSP in the buffer and log layer of wall-bounded
turbulence using resolvent analysis. The study also supports the
hypothesis that the SSP in the buffer and log layer are similar by
showing the similarities in the form of the principal forcing mode as
well as the structures of the flow correlations conditioned to high
forcing-intensity events that lead to the generation of the principal
forcing mode. 

\vspace{0.3cm}
This work was funded in part by the Coturb program of the European
Research Council (ERC-2014.AdG-669505) during the 2017 Coturb
Turbulence Summer Workshop at the UPM. B. J. M. is grateful for the 
support of ONR under N00014-17-1-3022. A.L.-D. acknowledges the
support of NASA under NNX15AU93A and of ONR under 
N00014-17-1-2310. The authors thank Prof. Javier Jim\'enez, Dr.
Yongseok Kwon, and Dr. Anna Guseva for their insightful comments.

Declaration of Interests: The authors report no conflict of interest.

\bibliographystyle{jfm}
\bibliography{nonlinear_bae}

\end{document}